# Defect Induced Room Temperature Ferromagnetism in Methylammonium Lead Iodide Perovskite

Sayantan Sil,[†#] Homnath Luitel,[‡#] Mahuya Chakrabarty,[$] Partha P. Ray,[‡] Joydeep Dhar,*[ζ] Bilwadal Bandyopadhyay,*[£] Dirtha Sanyal*[‡, ξ]

[†]Department of Physics, Jadavpur University, Kolkata 700032, India
[‡]Variable Energy Cyclotron Centre, 1/AF, Bidhannagar, Kolkata 700064, India
[ξ]Homi Bhabha National Institute, Training School Complex, Anushakti Nagar, Mumbai 400094, India
[$]Department of Physics, Basirhat College, Basirhat 743412, India
[ζ]Department of Chemistry, Birla Institute of Technology, Mesra, Ranchi 835215, India
[£]Saha Institute of Nuclear Physics, 1/AF Bidhannagar, Kolkata 700064, India

[#]Contributed equally

Corresponding Author. *E-mail: dirtha@vecc.gov.in

## Abstract

The defect tolerance nature of organic-inorganic hybrid perovskite is reflected from its stupendous growth in photovoltaic performances. The presence of lattice defect can manipulate or even gives rise to some exceptional properties which otherwise would have remained unseen. One of such properties reported in this article is the experimental observation of defect mediated room temperature ferromagnetism in methylammonium lead halide perovskite for the very first time, ably supported by ab-initio calculations. Theoretical analysis predicts the ferromagnetism principally arises from the iodide vacancies in the orthorhombic and cubic crystal phases but not in the tetragonal phase. The low temperature (100 K) ferromagnetic hysteresis loop was stable even at a high temperature of 380 K substantiating the fact that the origin of magnetism embedded in its defective nature.

**Introduction**

The rapid surge in the power conversion efficiency (PCE) of organic-inorganic hybrid lead halide perovskite based solar cell has generated huge excitement in the scientific community to achieve the maximum efficiency determined from the Shockley-Queisser calculations.[1-3] Exceptional photophysical[4-6] and charge transport properties[7-9] of organo lead halide perovskites have acted as a source of motivation, not only in exploring the field of photovoltaics, but to try for its multitude of applications in light emitting diodes (LEDs),[10] light emitting field effect transistors (LEFETs),[11] piezo-electric nanogenerators,[12-13] and non-volatile memory devices;[14] thus making hybrid perovskite as one of the highly followed research areas for material scientists in present days.

In spite of extraordinary rise in the photovoltaic performance of methylammonium lead iodide (MAPbI$_3$) perovskite which is a direct band gap semiconductor with a band gap of 1.6 eV,[15] the instability of organo lead halide perovskite in ambient has led to an extensive search to find out the origin of degradation and the ways to improve the atmospheric stability through rigorous theoretical and experimental analysis.[16-17] This resulted in the identification of lattice defect mediated degradation pathway in hybrid perovskites,[18-19] which further expedite with the exposure to light[20] and electric field.[21] As believed by different research groups, the frequently observed hysteresis in current density-voltage (*J-V*) plot during photovoltaic measurement is a consequence of lattice defects,[22-23] suggesting organo lead halide perovskites are sensitive towards easy defect formation. At room temperature, the electronic charge carriers in a field-effect transistor (FET) device are largely screen out by mobile ions as observed by Chin *et al*.[11] Our recent findings also substantiate earlier observations concerning defect mediated ionic transport in methylammonium lead iodide (MAPbI$_3$) perovskite and we were able to determine the identity of the vacancy sites using positron annihilation spectroscopy (PAS).[24-25] The organic methylammonium (MA$^+$) cation and iodide (I$^-$) anion, both contribute for defect formation,[24-25] but in some rare cases high energy lead (Pb$^{2+}$) vacancy can also form in the crystal lattice.[12] Thus, we observe that the lattice defects significantly influence the ionic polarization which in turn determines the magnitude of piezo-response in MAPbI$_3$.[12] Of late, the defective nature of MAPbI$_3$ is also successfully utilized in fabricating non-volatile memory devices.[14, 26-27] Therefore, the recent findings distinctly corroborate the pivotal role played by the lattice defects in governing physical properties of organo lead halide perovskites.

In last two decades scientists were searching new materials which could show ferromagnetism and semiconducting properties simultaneously for its possible application in spintronic devices. There were extensive research works on ZnO based spintronics material after theoretical prediction of room temperature ferromagnetism in ZnO and GaN.[28] Doping transition metal ions below 4 atomic percent in ZnO could induce ferromagnetism with a Curie temperature above room temperature. Intrinsically ZnO is n-type semiconductor. After doping with transition metal ions it can be converted into a p-type semiconductor, but making p-type ZnO is still an open challenge. It is very interesting that a new type of ferromagnetic ordering has been experimentally as well as theoretically observed, viz., defect induced ferromagnetism, in the diamagnetic materials. In last several years, a number of oxides e.g., ZnO, MgO, $TiO_2$, $BaTiO_3$ etc., showed ferromagnetic behaviour even at room temperature with the formation of atomic defects.[29-33] Ab-initio calculations in the frame work of density functional theory (DFT) have nicely explained the origin of such unusual ferromagnetism.[34-35]

**Results and Discussion**

In the present work, we report the first experimental observation of room temperature ferromagnetism in $MAPbI_3$, the organic-inorganic hybrid lead halide perovskite which is investigated extensively for its optoelectronic applications. The detailed theoretical analysis employing DFT calculations suggests the origin of ferromagnetic ordering in this material is again due to atomic defects, rendering ferromagnetism in a material which is otherwise a diamagnetic solid. The ab-initio calculations carried out on the three crystal structures of $MAPbI_3$ having all possible point defects aptly pinpoints the reason behind ferromagnetism without any ambiguity. Previously, Horváth and coworkers fabricated optically tuned magnetic switches with the extrinsically doped $MAPbI_3$ after substituting $Pb^{2+}$ with $Mn^{2+}$.[36] But, our observation about intrinsic ferromagnetic behaviour for a direct band gap semiconductor like $MAPbI_3$ is a real boost for envisaging its possible application in the field of spintronics. Furthermore, appearance of defect dependent ferroelectricity as well ferromagnetism in $MAPbI_3$ probably opens up a new dimension in the area of multiferroics for spintronic application.

The experimental evidence in support of the existence of defect driven magnetic moment for the polycrystalline $MAPbI_3$ sample having significant amount of point defects ($V'_{MA}$, $V''_{Pb}$, $V^{\bullet}_{I}$)[12] was performed in SQUID covering all the three crystal phases namely, orthorhombic, tetragonal and cubic ranging from 100 K to 380 K. The nature of defects was previously determined by PAS for the polycrystalline samples prepared following the reported synthetic procedure.[12] FIG. 1a represents the M-H curve for MAPbI$_3$ at different sample temperatures. A narrow but stable hysteresis loop with a constant coercive field and remnant magnetization over a broad temperature range (100 K to 380 K) signifies a Curie temperature well excess of room temperature. The temperature dependence of maximum saturated magnetization shown in FIG. 1b reflects the gradually decreasing trend in magnetic moment with increasing temperature, but only by 25% even after reaching a temperature (380 K) which is well above the room temperature. This suggests that the origin of ferromagnetism in MAPbI$_3$ must be related with the lattice defects, because defect density progressively increases as the temperature is raised.[24] We had to limit the maximum experimental temperature to 380 K beyond which MAPbI$_3$ degrades rapidly.[37-38] Two fairly steady dips in magnetic moment in the orthorhombic (below 160 K) and cubic (above 330 K) phases were connected through a rather flat magnetization domain in the tetragonal (160-330 K) phase indicating change in crystal structure is in some manner related with the origin of ferromagnetism.

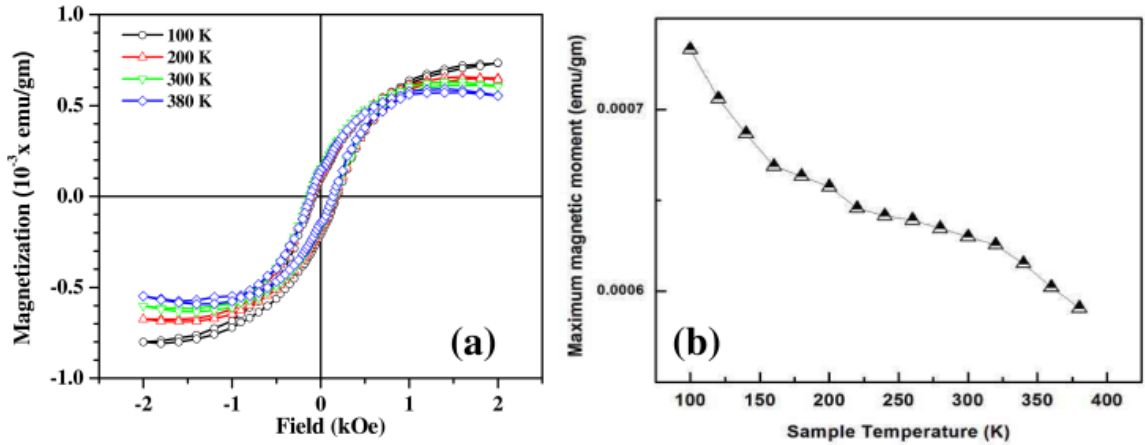

FIG. 1: (a) Magnetization hysteresis (M-H) loop for MAPbI$_3$ at different sample temperatures and (b) variation in maximum magnetic moment with respect to temperature.

In order to understand the magnetic properties in detail, entire spin polarized density of states for all the three different crystal structures of MAPbI$_3$ were studied in the MedeA software package build with VASP code[39-42] with the pseudo potentials, PAW, combining with generalized gradient approximation (GGA) and Perdew-Burke-Ernzerhof (PBE) exchange and correlation methods.[43] Three different unit cell structures of MAPbI$_3$; orthorhombic (Pnma symmetry, a=

8.84, b= 12.59, c= 8.56, α = β = γ = 90°), tetragonal (I4/mcm symmetry, a = b = 8.89, c =12.63, α = β = γ = 90°) and cubic (Pm-3m symmetry, a = b = c = 6.44 Å, α = β = γ = 90°) structures were optimized. The spin polarized density of states (DOS) calculations carried out on the three geometrically optimized structures in pristine (without defect) as well as for the samples having three different kinds of atomic vacancies (one vacancy in the system at a time viz. $V^{\bullet}_I$, $V''_{Pb}$ and $V'_{MA}$) to obtain theoretically effective magnetic moment. The spin-spin interaction studies have also been carried out on the system with non-zero magnetic moment by creating two atomic vacancies. The minimization of free energy determines whether parallel spin (ferromagnetic) or anti-parallel spin (anti-ferromagnetic) states are ground state.

After generating a single vacancy in a 2×2×2 super-cell, the corresponding magnetic moment have been calculated for all three crystal structures and tabulated in Table 1. The defect formation energies ($E_{df}$) have been calculated using the formula shown in Equation 1.[44-45]

$$E_{df} = E(MAPbI_3)_{\_D} - E(MAPbI3)_{\_P} - \sum_i n_i \mu_i + q[E_F + E_v + \nabla V] \quad (1)$$

where, $E(MAPbI_3)_{\_D}$ and $E(MAPbI_3)_{\_P}$ are the ground state free energy of defect induced system and the pristine system; $n_i$ is the number of atoms of type i (host/impurity atoms) that have been added ($n_i > 0$) or removed ($n_i < 0$) from the super cell. $\mu_i$ is the chemical potential of corresponding atom; q is the charge state, $E_F$ is the Fermi energy level and $\Delta V$ is the correction term.

As previously reported, in all the three crystal phases, the defect formation energy for iodine vacancy, $V^{\bullet}_I$ is the lowest while it is maximum for creating a lead vacancy, $V''_{Pb}$ due to its higher charge. Thus, it can be easily understood that the formation of $V^{\bullet}_I$ is energetically less demanding and more probable defect in $MAPbI_3$ system. It is interesting that the band gap of $MAPbI_3$ remains almost constant irrespective of the introduction or nature of the defects, while the position of the Fermi energy changes. This is the reason behind the "defect tolerance" nature of $MAPbI_3$.[46] In case of $MAPbI_3$ unit cell with $V'_{MA}$ and $V''_{Pb}$ the Fermi energy shifts towards valence band, which makes the system a p-type semiconductor while $V^{\bullet}_I$ makes the system n-type semiconductor as the Fermi level shifts approximately by an amount 1.6 eV towards the conduction band as compared to the pristine one. This is found to be true in all the three crystal phases of $MAPbI_3$. The band gap in the tetragonal state predicted from the theoretical calculation is higher as compared to other two crystal states and nicely follows the experimental value.[15] The total DOS for the pristine system in all three crystal structures are shown in FIG. 2.

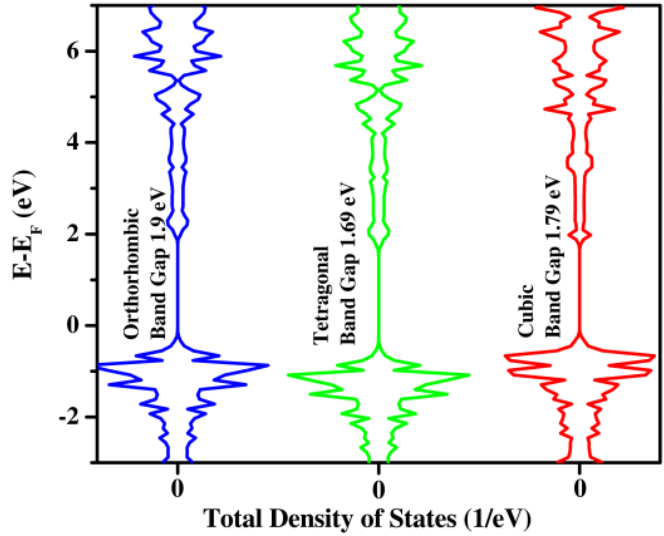

FIG. 2: The spin polarized total density of states for pristine $MAPbI_3$ in three different crystal structures.

Table 1. Magnetic moment in $MAPbI_3$ in pristine and after inducing defects in its three crystallographic phases

| Crystal Structure | Magnetic Moment Due to Defects ($\mu_B$) | | | Magnetic Moment without Defect (Pristine) ($\mu_B$) |
|---|---|---|---|---|
| | $V'_{MA}$ | $V''_{Pb}$ | $V^{\cdot}_{I}$ | |
| Orthorhombic | 0.00 | 0.00 | 0.74 | 0.00 |
| Tetragonal | 0.00 | 0.00 | 0.00 | 0.00 |
| Cubic | 0.00 | 0.73 | 0.68 | 0.00 |

From Table 1, it can be seen that one can introduce magnetic moment in $MAPbI_3$ system by incorporating defects. It is interesting that in the orthorhombic structure $V^{\cdot}_{I}$ can induce a magnetic moment, but other two types of defect cannot. Generation of one $V^{\cdot}_{I}$ induces magnetic moment of ~ 0.7 $\mu_B$ in the system. The primary source of magnetization is the presence of Pb-atom in the vicinity of the defective site. Around room temperature region (tetragonal state) none of the defects can induce any magnetic moment in the system. But, interestingly in the cubic structure both $V''_{Pb}$ and $V^·_I$ produce magnetic moment of ~ 0.7 $\mu_B$. The primary source of magnetism in $V''_{Pb}$ system is the iodine atom whereas in case of $V^·_I$ system, the Pb-atom is present in the vicinity of the defective site.

The spin polarized DOS (both partial and total) has been analyzed for all the systems to have a clear idea about the origin of defect induced magnetization. As shown in Figure 2, the DOS for pristine systems are symmetric in nature, suggesting an equal numbers of up and down-spin in electronic distribution. Thus, the pristine systems yield no net magnetic moment. However, in case of MAPbI$_3$ with cubic crystal structure, $V^·_I$, and $V''_{Pb}$ have asymmetric density of states (DOS) distribution as shown in FIG. 3.

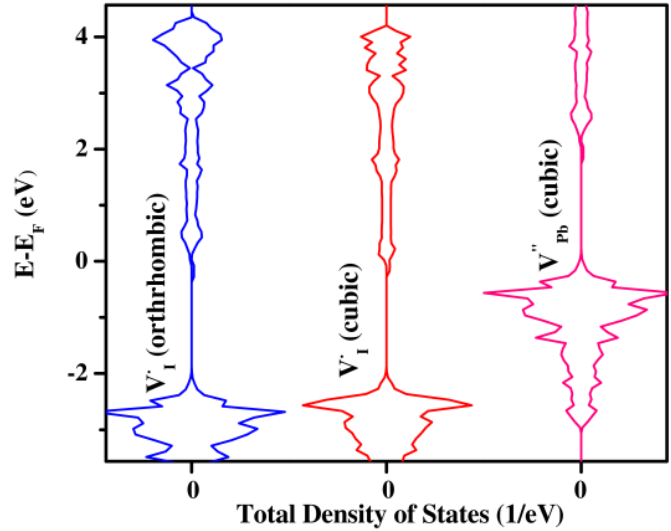

FIG. 3: Spin polarized total density of states for MAPbI$_3$ with $V^·_I$ and $V''_{Pb}$ in orthorhombic and cubic crystal structures having asymmetry in distribution.

The asymmetry in the total DOS can be observed from the partial DOS distribution in each case. In case of $V^·_I$ and $V''_{Pb}$, the asymmetry in DOS comes from the p-orbital electrons of Pb atoms and p-orbital electrons of I atom respectively. In case of $I_{inst}$ the asymmetry comes from I atom itself, present in the system. In all other non-magnetic cases, it has been found that the density distribution is symmetric as in the case of pristine system.

Spin-spin interaction studies have been carried out in the magnetic system to identify the nature of the ground state. Two identical defects have been created in the system and the ground state free energy has been calculated for parallel spin (ferromagnetic) as well as anti-parallel spin (anti-ferromagnetic) ordering. The cubic crystal system with two $V^·_I$ in the system produces 1.47

µ$_B$ magnetic moments. Also, the ferromagnetic ground state energy is less than its anti-ferromagnetic state, which clearly indicates that the ferromagnetic state is more stable in cubic MAPbI$_3$ system having V$^·_I$. In case of V$''_{Pb}$ in a cubic system, the spin-spin interaction study suggests that the anti-ferromagnetic coupling is more favored than its ferromagnetic counterpart.

In summary, we have demonstrated experimental evidence in support of room temperature ferromagnetism for MAPbI$_3$. Ab-initio calculations in the framework of density functional theory predict that ferromagnetism arises due to presence of iodine vacancy, but only for orthorhombic and cubic crystal structures. For tetragonal phase iodine vacancy is unable to induce any magnetic moment. But experimentally we observed the existence in magnetization even in the temperature region where mainly tetragonal phase is predominant. The possible reason may be the coexistence of two structures at a particular temperature.

## Acknowledgements

S. S acknowledges University Grants Commission (UGC) for providing NET-Senior Research Fellowship.